# Role of charged defects on the electrical and electro-mechanical properties of rhombohedral Pb(Zr,Ti)O$_3$ with oxygen octahedra tilts

Tadej Rojac,[1] Silvo Drnovsek[1], Andreja Bencan[1], Barbara Malic[1], and Dragan Damjanovic[2]

[1]Electronic Ceramics Department, Jožef Stefan Institute, Jamova cesta 39, 1000 Ljubljana, Slovenia

[2]Ceramics Laboratory, Swiss Federal Institute of Technology in Lausanne – EPFL, 1015 Lausanne, Switzerland

**Abstract**

Oxygen octahedra tilting is a common structural phenomenon in perovskites and has been subject of intensive studies, particularly in rhombohedral Pb(Zr,Ti)O$_3$ (PZT). Early reports suggest that the tilted octahedra may strongly affect the domain switching behavior, dielectric and piezoelectric properties of PZT ceramics. In a way similar to that proposed for tilts, however, charged defects, associated with oxygen vacancies, may also inhibit the motion of the domain walls, resulting macroscopically in pinched hysteresis loops and reduced piezoelectric response. Here, we revisit the early studies on rhombohedral PZT ceramics with tilted octahedra by considering a possible role of both tilts and charged defects on domain-wall motion. We show that the observed pinched hysteresis loops are likely associated with the presence of defect complexes containing charged oxygen vacancies, and not tilts as suggested in some cases. Regardless of the presence or absence of long-range ordered tilts in rhombohedral PZT, the effect of charged defects is also prominent in weak-field permittivity and piezoelectric properties, particularly at sub-Hz driving conditions where the conductivity, related to the motion of oxygen vacancies, gives rise to strong frequency dispersion.

## I. INTRODUCTION

The macroscopic electrical and electromechanical responses under weak (subswitching) and strong (switching) driving fields of the most widely used perovskite ceramics, Pb(Zr,Ti)O$_3$ (PZT), are affected by various structural and microstructural features, encompassing atomic, nano and microscopic length scales; these features are i) charged atomic defects, such as oxygen vacancies [1–6], ii) tilting of the oxygen octahedra in the perovskite lattice [7–10], iii) crystal structure, including the ferroelectric distortion of the unit cell and the proximity of a phase boundary at which two crystal structures coexist [1,4,11,12], iv) ferroelectric domain structure and grain size [12–14], and v) elastic coupling between grains [15,16].

Tilting (or rotation) of oxygen octahedra is a common phenomenon in compounds and solid solutions consisting of the perovskite ABO$_3$ structure [17]. Within a comprehensive study on several complex perovskites, an empirical relationship has been established between the octahedra tilting and the perovskite tolerance factor ($t$) according to which the onset of tilting is likely to occur in perovskites whose tolerance factor $t$<0.985 [18]. This general principle can be applied to the PbTiO$_3$–PbZrO$_3$ phase diagram and explains the onset of tilting at room temperature in rhombohedral PbZr$_x$Ti$_{1-x}$O$_3$, which occurs at composition $x$~0.6 [19]. In the past twenty years, a number of investigations have been performed on the relationship between octahedra tilts and macroscopic properties of PZT ceramics [7–10,20,21]. These initial studies on macroscopic behavior have recently been extended to those on a local level, revealing the unexpected role of the tilts in interfaces, such as domain walls [22,23].

It has been suggested that the polarization switching in undoped, rhombohedral PZT ceramics is strongly affected by the presence of octahedra tilts [7,8,20]. The pinched or double-like polarization–electric-field (*P-E*) hysteresis loops observed in undoped PZT ceramics within the compositional range 85/15 > Zr/Ti > 55/45 and, in particular, within the *R3c* phase region, which is characterized by long-range tilting of oxygen octahedra, was explained in terms of pinning of domain walls by oxygen octahedra rotations occurring through elastic interactions. Studies on the temperature dependence of *P-E* loops seemingly confirmed this point. In addition, the authors report on the inability to de-age the samples (i.e., to de-pinch the hysteresis loops) by quenching or by sintering the ceramics in PbO-rich atmosphere.

The influence of the tilt patterns on weak-field dielectric and piezoelectric properties of undoped PZT was later published by Eitel and Randall [10]. They reported on a reduction of the longitudinal converse piezoelectric coefficient and dielectric permittivity in the *R3c* phase relative to the *R3m* phase. On the basis of a transmission electron microscopy (TEM) analysis, it was proposed that the suppression of domain-wall motion in the tilted *R3c* phase is due to the pinning effect of the anti-phase boundaries, which are created within the grains due to the inversion of the rotational sense of the tilts. A similar decrease in the piezoelectric coefficient was observed at the onset of the appearance of long-range ordered tilts in Sr- and Nb-co-doped morphotropic PZT [9,21].

A possible role of octahedra tilts in domain-wall pinning has been also framed within a recent theoretical study by Beanland [24]. It has been predicted that the condition of continuity of octahedra tilts across the domain wall necessarily leads to different octahedra tilting (different local tilt system) at the wall, revealing a unique structural coupling between tilts and domain walls.

Even though the macroscopic behavior of rhombohedral PZT ceramics at switching and subswitching fields appears consistent with the proposed effects of octahedra tilts, one should be particularly careful in deriving such conclusions. This is because the domain switching and piezoelectric behavior of undoped PZT may be affected by domain-wall pinning centers other than tilts, for example, by oxygen vacancies [1,6,25], which can be created as a result of PbO sublimation during sintering [26–28]. In fact, hysteresis loop pinching is most often attributed to interactions between point defect complexes and domain walls [2]. However, the origin(s) of pinching may still be multiple and rather complex. For example, some compositions in the $(Na_{0.5}Bi_{0.5})TiO_3$–$BaTiO_3$ solid solution exhibit pinched *P-E* loops with no clear evidences that the origin of the loop pinching is point defect complexes and it is most likely structural in character [29,30]. The separation of the pinning by the tilts and other effects from that of the charged defects is thus non-trivial and requires a systematic analysis of the aging and de-aging behavior of PZT.

In this paper, we revisit the switching behavior and weak-field properties of PZT by considering both charged defects and tilts as domain-wall pinning centers. By comparing the behavior of *R3c* PZT75/25 and *R3m* PZT60/40, we show that the mechanism responsible for the pinched hysteresis loops is dominated by charged defects, most probably lead-vacancy–oxygen-vacancy ($V_{Pb}''$–$V_O^{\bullet\bullet}$) defect complexes, and not octahedra tilts as earlier suggested for this system. While the pinning effect of the tilts, consistent with earlier studies, is observed in the weak-field frequency dispersion of permittivity and piezoelectric $d_{33}$ coefficient in the frequency range from 1 to 100 Hz, a dispersion in these properties is also revealed at low driving frequencies (in the range from 0.01 to 1 Hz), and can be attributed to conductivity-related mechanisms involving oxygen-vacancy motion. The results thus reveal an important role of charged defects in undoped PZT, which may be extended to other perovskites and which must be considered in addition to the long-range structural distortion associated with the octahedra tilts.

**II. EXPERIMENTAL DETAILS**

$PbZr_xTi_{1-x}O_3$ ceramics with compositions $x$=0.75 and $x$=0.6 were prepared by conventional solid state reaction from starting powders PbO (Sigma-Aldrich, 99.9%), $ZrO_2$ (Tosoh, 99.1%) and $TiO_2$ (Alfa-Aesar, 99.8%). 1 mol% excess of PbO was added in both compositions. A doped sample was prepared by substituting 1 mol% of $ZrO_2/TiO_2$ with $Nb_2O_5$ (Sigma, 99.9%) according to the nominal formula $Pb(Zr_{0.75}Ti_{0.25})_{0.99}Nb_{0.01}O_3$. No PbO excess was added to this composition. The three ceramic compositions are further denoted as PZT75/25, PZT60/40 and PZT75/25-Nb. The expressions "non-tilted" and "tilted" PZT refer, respectively, to the absence or presence of long-range ordered oxygen octahedra tilts in the structure, which were determined by identifying the space group symmetry: *R3m* for "non-tilted" and *R3c* for "tilted" PZT (see Section IIIA for details on structural identification of PZT).

The three PZT powder batches were prepared following the same procedure. Approximately 400 g of the starting powder mixture were first homogenized in a Retsch PM400 planetary mill for 2 h at 200 $min^{-1}$ of main disc frequency. We used a 500 mL yttria stabilized zirconia (YSZ) milling vial, 100 YSZ milling balls with diameters of 10 mm and 200 mL of isopropanol as dispersion medium. After mixing, the powder mixtures were dried at 95°C and then calcined at 900°C for 2 h in a closed alumina crucible using 5 °C/min of heating and cooling rates. The calcined powders were attrition milled (Netzsch attritor mill) in isopropanol at 800 $min^{-1}$ of rotational frequency for 4 h using a 500 mL YSZ milling vial filled with 1000 g of YSZ milling balls with diameters of 3 mm. This was followed by drying the powders at 95°C.

The milled and calcined powders were compacted into cylindrically shaped pellets by uniaxial pressing at 50 MPa, followed by isostatic pressing at 300 MPa. The pellets were then sintered in a closed alumina crucible at 1250°C for 2 h with 5 °C/min of heating and cooling rates. The pellets were sintered in air and were covered with a packing powder with the same composition as the pellet; the only exception is the ceramics used for the quenching experiments with different cooling rates (see Section IIIA), in which case the pellet was sintered in $O_2$ atmosphere and was surrounded by a packing powder consisting of $PbZrO_3$ with an addition of 2 mol% of PbO.

The density of the PZT ceramics was determined geometrically. For the calculation of the relative density we used the theoretical density of 8.03 g/cm$^3$ for PZT75/25 and PZT75/25-Nb (ICSD #202845), and 8.04 g/cm$^3$ for PZT60/40 (ICSD #77595).

The phase composition of the ceramics was analyzed by X-ray powder diffraction (XRD) using a PANalytical X'Pert Pro diffractometer with CuKα1 radiation. Room-temperature measurements were done in a 2-theta range 10–70° with a step 0.034° and scan time 100 s. For the identification of the space group symmetry of PZT, the samples were measured in a 2-theta range 35.6–36.9° with a step of 0.017° and scan time 12000 s. The XRD analyses were performed either on powders, prepared from crushed sintered pellets, or directly on sintered pellets. "In-situ" high-temperature XRD analysis was performed using the same diffractometer in a high-temperature configuration with CuKα1/Kα2 radiation and an Anton Paar HTK-1200 oven. The XRD patterns were collected in a 2-theta range 35.6–37.1° with a step of 0.013° and scan time 15000 s. Simulations of XRD patterns of the low-temperature *R3c* PZT75/25 (ICSD #202845) and high-temperature *R3m* PZT75/25 (ICSD #202846) were performed using a Topas R software package (Bruker AXS, Germany).

The microstructure, domain structure, phase composition and compositional homogeneity of the ceramics were analyzed using a scanning electron microscope (SEM) JSM-7600 F equipped with a JEOL energy-dispersive X-ray spectrometer (EDXS). EDXS analyses were performed at 15 keV. The preparation of the samples' surfaces for SEM followed a standard metallographic procedure. The grain size was estimated from SEM images using the standard intercept method (ASTM Standard E112-13).

According to XRD analysis and backscattered-electron (BE) SEM imaging (not shown), the PZT75/25, PZT60/40 and PZT75/25-Nb ceramics were phase-pure perovskites.

The relative density and grain size were, respectively, 96% and 8.4 μm for PZT75/25, 96% and 11.0 μm for PZT60/40, and 97% and 6.8 μm for PZT75/25-Nb ceramics. We note that the grain sizes are above the critical grain size of ~2 μm, below which the piezoelectric and dielectric properties of PZT become significantly affected [14]. A similar domain structure was observed for the three PZT compositions and the compositional homogeneity was confirmed by performing statistical EDXS point analyses on ~20 arbitrarily selected spots on the sample's surface (not shown).

Quenching experiments were performed first by rapidly immersing a PZT pellet of typical dimensions 6-7 mm in diameter and 0.5 mm in thickness into a tube furnace, pre-heated at 600°C. After exposing the sample for 5 min at this temperature, the pellet was quenched either by dropping it into a 70°C-pre-heated water ("water-quenched") or by pulling it out from the furnace, in which case the pellet was rapidly cooled in air ("air-quenched"). No evidences of cracks and no changes in the dimensions and mass of the pellets were observed after quenching; however, a clear colour brightening of the pellets occurred as a result of quenching.

For electrical and electromechanical characterizations, the sintered pellets were first cut and thinned either to 0.5 mm for switching experiments (section IIIA) or 4 mm for permittivity, direct and converse piezoelectric measurements (section IIIB). After polishing the two surfaces of the pellets, all the samples (except the quenched ones) were additionally heated at 600°C for 1 h with a heating and cooling rate of 5 °C/min and 1 °C/min, respectively; this was done to release mechanical stresses induced by polishing. The pellets were finally electroded with Au/Cr by sputtering. For permittivity and piezoelectric measurements, the PZT75/25 and PZT60/40 samples were poled at room temperature in two successive steps: first with a DC field of 4 kV/mm for 10 minutes and, second, with 5 kV/mm for another 10 minutes. The PZT75/25-Nb sample was poled in one step at 3kV/mm DC field for 10 minutes at room temperature.

*P-E* loops were measured using an aixACCT TF 2000 analyzer. The measurements were performed by applying to the samples electric fields with incremental amplitudes of 0.1, 1, 2 and 3 kV/mm at 10 Hz. Two sinusoidal waveforms were applied at each field step. During the measurements the samples were immersed in silicone oil.

Permittivity measurements were performed with the aixACCT TF 2000 analyzer by applying to the samples bipolar electric fields of amplitude in the range 0.05–0.4 kV/mm and frequency 0.01–100 Hz. For each field of selected amplitude and frequency, we applied three sinusoidal waveforms in sequence. The results are presented in terms of the real part of the complex permittivity ($\varepsilon'$) and dielectric losses ($tan\delta_d = \varepsilon''/\varepsilon'$ where $\varepsilon'$ is the real part and $\varepsilon''$ is the imaginary part of the complex permittivity).

The direct piezoelectric $d_{33}$ response was measured using a dynamic press [31], as a function of alternating stress amplitude in the range 0.5–2 MPa and frequency in the range 0.01–100 Hz. Before measurements, a static pre-stress of 4.7 MPa was applied to the samples. After each measurement, the samples were checked for possible depoling; in all the cases, the samples showed the same $d_{33}$ before and after the measurements. The results are presented in terms of $d_{33} = Q_0/F_0$ where $Q_0$ and $F_0$ are amplitudes of measured charge and applied force, respectively, and $tan\delta_p = d''/d'$ where $d'$ and $d''$ are real and imaginary components of the piezoelectric coefficient, respectively.

The converse piezoelectric $d_{33}$ response was measured with a home-made setup using a fiber-optic displacement sensor (MTI 2100 Fotonic Sensor) as reported in Ref. [32]. During measurements, the samples were exposed to continuous bipolar electric-field cycling with amplitudes in the range 0.05–0.4 kV/mm and frequency in the range 0.06–100 Hz. The data were collected with the oscilloscope and the lock-in amplifier. The results are represented in terms of mechanical-displacement–electric-field ($\Delta l$-$E$) hysteresis loops or in terms of $d_{33}$ as described above for the direct piezoelectric effect. The hysteresis-loop de-pinching, induced by continuous bipolar electric-field cycling, was followed by measuring the phase angle of the third harmonic signal of the displacement response ($\delta_3$) using a lock-in amplifier, similarly as reported in Ref. [33].

The analysis of the $P$-$E$, charge-density–stress ($D$-$\sigma$) and $\Delta l$-$E$ hysteresis loops, measured at subswitching fields, was performed by determining the area of the loops, $A$, either numerically or using the equation:

$A = \pi F_0 R_0 \sin\delta$ (1)

where $F_0$ is the driving-field (electric or stress) amplitude, $R_0$ is the amplitude of the material's response (charge or mechanical displacement) and $\delta$ is the phase angle between the material's response and driving field. The area of the hysteresis was determined at different field amplitudes and the slope of $\log(A)$ versus $\log(F_0)$ plots was determined by fitting the data points with a linear function. Details about this analysis are reported in Ref. [34].

## III. RESULTS AND DISCUSSION

### A. Domain switching behavior

The $P$-$E$ loops of the as-sintered PZT75/25 ceramics with tilted $R3c$ structure and PZT60/40 ceramics with non-tilted $R3m$ structure are shown in Fig. 1 (solid curves). PZT75/25 shows stronger pinching (Fig. 1a, solid curve) and, thus, a lower remanent polarization ($2P_r$=5.0 μC/cm$^2$) and maximum polarization ($P_m$=20 μC/cm$^2$) as compared to PZT60/40 ceramics (Fig. 1b, solid curve; $2P_r$=30 μC/cm$^2$, $P_m$=27 μC/cm$^2$). Similar pinched loops were observed earlier in both *R3c* and *R3m* PZT [7,8,20]. In those studies, it was suggested that the pinching behavior originates in the dynamic interaction between the polarization reversal and the oxygen octahedra tilts during the switching process. A qualitative model was proposed, which assumes that, while the spontaneous polarization switches under the electric field, the inversion of the rotational sense of the octahedra tilts does not follow the polarization reversal, resulting in internal stresses in the grains. These stresses produce a restoring force that tends to re-establish the initial domain configuration, giving rise, macroscopically, to pinched *P-E* loops. The proposed explanation was supported by the observed stronger loop pinching and reduced maximum measured polarization for PZT compositions lying in the octahedrally tilted *R3c* phase region or close to the *R3c-R3m* phase boundary. Apparently, the proposed model is consistent with our results (Fig. 1, compare solid curves).

In addition to the octahedra tilts, other pinning centers should be considered, e.g., charged defects associated with oxygen vacancies ($V_O^{\bullet\bullet}$). Pinched loops, like those shown in Fig. 1 (solid curves), are commonly observed in acceptor-doped, so-called "hard" ferroelectric materials, such as Mn- and Fe-doped PZT [2,33], and were also observed in undoped perovskites [34–37], including PZT [34]. In those cases, the "hardening" behavior and pinched hysteresis loops are attributed to the presence of defect complexes of the type acceptor-site–$V_O^{\bullet\bullet}$. Here, the acceptor site refers either to intentionally added dopant (e.g., $Fe_{Zr,Ti}'$ in Fe-doped PZT [6]), impurity (e.g., $Fe_{Ti}'$ in BaTiO$_3$ where Fe$^{3+}$ is an acceptor impurity [35]) or defects created during high-temperature ceramic processing (e.g., $V_{Pb}''$ in undoped PZT, formed as a result of PbO loss [4,25]). Through diffusion of $V_O^{\bullet\bullet}$, such defect complexes may arrange with time into an ordered, equilibrium configuration by aligning along the direction of the spontaneous polarization within a domain, from where they stabilize (or pin) the domain walls [2,3,6,35,38,39]. This aging process is thus diffusion controlled and starts as soon as the ceramics enter the ferroelectric phase during cooling from the sintering temperature.

In order to discriminate between the effects of the tilted octahedra and defect complexes on the domain switching behavior of PZT, we performed quenching experiments. Quenching is a valuable methodology, particularly for studying the interactions between defect complexes and domain walls, aging effects, domain-wall dynamics and domain structure evolution in disordered ferroelectric materials [40–42].

One of the ways to de-age the sample or depin the domain walls from the defect complexes is to convert the ordered defects (aligned with polarization) into a disordered state. This can be done by annealing the ceramics at a temperature higher than the Curie temperature ($T_C$), i.e., in the paraelectric phase, where the complexes are disordered due to the absence of the spontaneous polarization. The annealing is then followed by a rapid cooling (quenching) of the ceramics to room temperature. Due to kinetic restrictions, the rapid cooling prevents the oxygen vacancies to diffuse from the disordered state back to their equilibrium ordered positions, so that the disordered state becomes frozen at room temperature. Such disordered defect state releases locally the domain walls, leading macroscopically to depinching (opening) of the *P-E* loop [34,43].

In contrast to point-defect complexes, we assume and, subsequently, confirm by XRD analysis, that the octahedra tilts will be re-ordered after the quenching through the tilt transition. This is because the ordering of the tilts requires only small displacements of the $O^{2-}$ ions, contrasting the rather time-dependent and slow $O^{2-}$ (or $V_O^{\bullet\bullet}$) diffusion, which is necessary for the ordering of the defect complexes. Therefore, we expect that the defect complexes will be frozen in their disordered state by quenching from above $T_C$, whereas the tilts will re-order once crossing the tilt transition ($T_t$=125°C for PZT75/25 [1]). While these arguments appear reasonable, the ordering behavior of the tilts upon quenching the PZT75/25 must be proven. This is because the tilt transition in rhombohedral PZT, particularly for compositions with Zr/Ti ratio <80/20, was found to be of a gradual order-disorder type [7,20,44–47]. This means that a transition is expected upon cooling from the disordered (local) tilts in the *R3m* phase to the ordered tilts in the *R3c* phase. The high cooling rate, provided by the quenching, might thus still affect the ordering behaviour of the tilts [48].

The *P-E* loops of PZT75/25 and PZT60/40 ceramics after quenching from 600°C into water is shown in Fig. 1 (dashed curves). A clear transition from a pinched loop to a depinched (open) loop is observed for both PZTs, suggesting that the strong loop pinching of the as-sintered ceramics (Fig. 1, solid curves) is caused by the pinning effect of the defect complexes. Within the experimental error, the remanent polarization of the two quenched ceramics is the same ($2P_r$=70 μC/cm$^2$; see also inset of Fig. 1). This means that, once the domain walls are released from the charged defects by quenching, the tilted *R3c* and non-tilted *R3m* PZT ceramics exhibit very similar switching behaviour. Thus, in contrast to what has been proposed earlier [7,8,20], our results rather suggest that the tilts could not be entirely responsible for the pinched loops in rhombohedral PZT.

It should be noted that the coercive field ($E_c$) and internal bias field ($E_{ib}$), i.e., the shift of the loop along the electric-field axis, appear slightly higher in quenched PZT75/25 than in quenched PZT60/40 (see inset of Fig. 1). According to the literature, $E_{ib}$ is usually attributed to a preferred orientation of the defect complexes in the ceramics along the poling direction, and may be understood as an additional coercive field arising due to these preferentially aligned defects [2,3,49]. Therefore, the larger $E_{ib}$ of the quenched PZT75/25, relative to that of PZT60/40, could be attributed to a higher concentration of $V_O^{\bullet\bullet}$ in these ceramics, which is also consistent with the more pronounced loop pinching of the as-sintered PZT75/25 as compared to that of the as-sintered PZT60/40 (compare solid-curve loops in Fig. 1a and 1b).

In the case of PZT60/40 ceramics, whose composition is close to that at the morphotropic phase boundary (MPB), we should not exclude a possible effect of the MPB on $E_{ib}$. For example, $E_{ib}$ in Fe-doped PZT was found to decrease by approaching the MPB [2].

We shall now verify whether the long-range ordering of the tilted octahedra was re-established after quenching the PZT75/25 sample. For this purpose, we used XRD analysis. The *R3m* and *R3c* space group symmetries of PZT can be distinguished through the appearance of additional low-intensity XRD peaks in the *R3c* PZT pattern, which are absent in the *R3m* phase. These peaks, shown in the simulated pattern in Fig. 2a (see arrows in the upper pattern), are present in the *R3c* phase because of its lower symmetry, relative to the *R3m*. The reduced symmetry of the *R3c* phase is related to the octahedra tilting: the glide plane "*c*" in the *R3c* symmetry replaces the mirror plane "*m*" of the *R3m* to represent the doubling of the unit cell. Hence, by considering this point, one can use the low-intensity peaks to identify indirectly the long-range ordering of the tilted octahedra in PZT. An experimental example is shown in Fig. 2b: the $(1\bar{3}\bar{1})$ peak of the *R3c* PZT75/25, observed at room temperature (see 25 °C (1)), clearly disappears in the *R3m* phase region at 200°C, and re-appears once the *R3c* phase is re-established upon cooling the sample to room temperature (see 25 °C (2)).

Figure 3 displays the XRD patterns of PZT ceramics of different compositions before (as-sintered) and after quenching, highlighting the 2-theta region in which the $(1\bar{3}\bar{1})$ peak of the *R3c* phase should appear. The absence of the $(1\bar{3}\bar{1})$ peak in the as-sintered PZT60/40 and its presence in the as-sintered PZT75/25 at room temperature confirm, respectively, the *R3m* and *R3c* symmetry of the two ceramics, as expected from the PbTiO$_3$–PbZrO$_3$ phase diagram [1]. In the case of the tilted *R3c* PZT75/25, the $(1\bar{3}\bar{1})$ peak is clearly visible immediately after quenching and after subsequent aging of the quenched sample. This confirms that the octahedra tilts were re-ordered on a long-range scale upon quenching from above 600°C. We emphasize that the PZT75/25 consists of ordered octahedra tilts both in the as-sintered and quenched state (Fig. 3), however, the polarization response is distinctly different (compare loops in Fig. 1a). We can thus safely assume that the tilts do no play a major role in the initially observed pinching of the loop (see Fig. 1a, solid curve).

Another possibility that we consider is the electric-field induced transition between the *R3c* and *R3m* phases during loop measurements. In-situ structural studies performed by Yang *et al.* [50] showed that strong electric fields (>10 kV/cm) can induce a transition from the *R3m* to the *R3c* phase in a doped PZT at temperatures in the vicinity of the tilt-transition temperature (within a temperature range ~10–20 K around the zero-field tilt transition temperature). Assuming that the pinning of domain walls in the tilted *R3c* phase is stronger than in the non-tilted *R3m* phase and that the electric field favors the *R3c* phase, we may expect that such field induced phase transition would lead to a stronger pinching of the loops at high electric fields, even in quenched samples. In contrast, the hysteresis loops of quenched samples show an open shape (Fig. 1), which is not consistent with the reported field induced *R3m*-to-*R3c* phase transition.

In order to investigate more closely the antiphase boundaries (APBs) associated with the octahedra tilts in *R3c* PZT, we also performed a TEM analysis. Since the APBs may act as pinning sites for domain walls [10,51] and considering that their formation could be, in principle, affected by the quenching, we analyzed the APBs in the as-sintered and quenched PZT75/25 using TEM. As for the appearance of the APBs within the grains, we could not identify any significant difference between the samples before and after quenching (not shown).

In the next step, we performed aging studies on the quenched PZT samples. After quenching and *P-E* hysteresis measurements (Fig. 1, dashed curves), the PZT75/25 and PZT60/40 ceramics were aged at room temperature without applying external field. To check for polarization changes, the *P-E* loops were re-measured after selected periods of aging within seven days. Both PZT75/25 and PZT60/40 showed similar aging behavior, i.e., a reduction of $2P_r$ and an increase in $E_{ib}$ with time (not shown). The similarity in the aging between the composition with tilted and non-tilted octahedra suggests a minor role of the tilts in the aging behavior and reinforces the role of defect complexes.

To further show that the defect complexes are mainly responsible for the pinched loops in the tilted and non-tilted PZT, we quenched the PZT75/25 samples from above $T_C$ in different media to provide different cooling rates. If we assume that the domain walls are pinned by the defect complexes, whose re-orientation requires diffusion and, thus, time, it is expected that the level of de-aging (followed by loop depinching) will depend on the cooling rate: the slower the cooling, the more time is provided for the complexes to re-arrange from the disordered state, achieved at $T>T_C$, back to the ordered state at $T<T_C$ [34]. This is seen in Fig. 4a; with respect to the sample rapidly quenched in water ($H_2O$-quenched), the sample cooled in air (air-quenched) shows a weaker polarization response with a slanted loop and a much smaller $2P_r$ (35 μC/cm$^2$ for air-quenched versus 65 μC/cm$^2$ for water-quenched sample). The progressive opening of the loop from the initial aged and pinched state (Fig. 4a, as-sintered) by increasing the cooling rate is consistent with the presence of defect complexes as domain-wall pinning centers and their order(aged)-disorder(de-aged) behavior.

Another way to depin the domain walls from the defect complexes, alternative to quenching, is to compositionally modify the PZT with a donor dopant, leading to the so-called "soft" PZT [1]. A classical donor dopant for PZT is $Nb_2O_5$, where $Nb^{5+}$ substitutes for the B-site $Zr^{4+}$ and $Ti^{4+}$ cations, resulting in donor sites, $Nb^{\bullet}_{Zr,Ti}$. The formation of these positively charged Nb donor sites is expected to decrease the concentration of positively charged $V_O^{\bullet\bullet}$, relative to undoped PZT, resulting in compensating negatively charged lead vacancies, $V_{Pb}''$ [1,5,25,52]. We note that the as-prepared Nb-doped PZT75/25 sample was identified with an *R3c* symmetry, evidenced by the presence of the $(1\bar{3}\bar{1})$ peak in the XRD pattern (Fig. 3). Therefore, the structure of these doped ceramics consists of long-range ordered octahedra tilts.

The *P-E* loop of Nb-doped PZT75/25 in the as-sintered state appears open with no signs of pinching (Fig. 4b). Unlike undoped PZT75/25 and PZT60/40, we observed very little aging in this doped sample. Both these observations are consistent with the expected low concentration or absence of domain-wall pinning centers ($V_O^{\bullet\bullet}$) as a result of donor doping [5]. It is thus reasonable to assume that $V_{Pb}''$ and $Nb^{\bullet}_{Zr,Ti}$ are created in PZT75/25-Nb, which are not likely to interact with the polarization in the material and, in principle, do not have a tendency to create orientable defect complexes [25]. Therefore, no evidence is found which would suggest that the octahedra tilts in the donor-doped sample have any substantial effect on the switching behavior in terms of loop pinching. We should emphasize, however, that the effect of the donor dopant on the switching should be taken with care as the mechanisms of "softening" in PZT are still not well understood [5,25]. The data are, however, consistent: either disordering the point defect complexes by quenching (Fig. 1 and 4a) or reducing their concentration by doping (Fig. 4b), lead to depinning of domain walls, even in samples, such as PZT75/25, whose structure consists of long-range ordered octahedra tilts.

In summary, the results give a number of indirect evidences on the presence of point-defect pinning centers in *R3c* and *R3m* PZT, most likely associated with $V_O^{\bullet\bullet}$, which play a dominant role in the switching behavior of the ceramics. In undoped PZT, the most probable and also the most often assumed, are the $V_{Pb}''$–$V_O^{\bullet\bullet}$ complexes, which could form as a result of PbO loss during sintering [1,6,25–28]. Recent theoretical studies on PbTiO$_3$ (PT) revealed that the $V_{Pb}''$–$V_O^{\bullet\bullet}$ complexes in nominally undoped PT and $Fe_{Ti}'$–$V_O^{\bullet\bullet}$ complexes in acceptor-, Fe-doped PT tend to behave very similarly [25]. In fact, it is expected that such complexes would align preferentially in the direction of spontaneous polarization within domains, leading to domain-wall clamping and macroscopic "hardening" behavior. This may explain why nominally undoped PZT behaves similarly as the acceptor-doped PZT in terms of pinched loops and aging. In addition, our data show that the macroscopic behaviour of *R3c* and *R3m* PZT, including pinched polarization loops (Fig. 1, solid-curve loops), aging and de-aging (Fig. 1, dashed-curve loops, and Fig. 4a), is similar to that of acceptor-doped PZT. Thus, considering these theoretical and experimental results, we can infer on the presence $V_{Pb}^{''}$–$V_O^{\bullet\bullet}$ complexes in undoped, rhombohedral PZT. Another possibility would be complexes created by unintentional incorporation of acceptor impurities, e.g., $Fe_2O_3$, in the perovskite [35,53].

**B. Hysteresis and frequency dependence of dielectric permittivity, and direct and converse longitudinal piezoelectric coefficient**

Figure 5 shows the frequency dependence of permittivity and direct piezoelectric $d_{33}$ coefficient of PZT75/25 and PZT60/40 ceramics. We begin by presenting the permittivity data, which is followed by the piezoelectric data.

Qualitatively, the same frequency dependence of permittivity is observed for the two PZT compositions (Fig. 5a, b). In the high frequency range, ~0.5–100 Hz, $\varepsilon_{33}{'}$ follows a $\log(1/\omega)$ dependence, where $\omega$ is frequency, as indicated by a dashed line in Fig. 5a,b. This logarithmic behavior is accompanied by a weak dispersion of $tan\delta_d$, which is between 0.1 and 0.2 for the two samples. The $\log(1/\omega)$ dispersion has been observed earlier in other ferroelectrics, including Nb-doped PZT and $Bi_4Ti_3O_{12}$ [54]. The $\log(1/\omega)$ behavior was assigned to domain-wall displacement processes, occurring in a random energy landscape. The irreversible domain-wall displacements are macroscopically reflected in the nonlinearity (dependency of the coefficient on the external driving-field amplitude) and nonlinear hysteresis [54–56], which were consistently observed in PZT75/25 and PZT60/40 (not shown).

In the low frequency range, below ~0.5 Hz, $\varepsilon_{33}'$ and $tan\delta_d$ of both PZT compositions (Fig. 5a,b) deviate significantly from the $\log(1/\omega)$ dependence and show a steeper frequency dispersion with $tan\delta_d$ approaching 1 ($\varepsilon''=\varepsilon'$) in the low frequency limit. An analogous dispersion was also observed in rhomobohedral (Fe-doped), "hard" PZT and was attributed to the presence of $V_O^{\bullet\bullet}$, whose hopping from one oxygen site to another under the field will affect $\varepsilon'$ and $\varepsilon''$ at low frequencies [5]. The similarities in the dielectric and piezoelectric responses of undoped and "hard" rhombohedral PZT were proposed to be due to the presence of $V_O^{\bullet\bullet}$ in the two compositions, that is, i) $V_{Pb}''-V_O^{\bullet\bullet}$ complexes in undoped PZT, resulting from the PbO loss, or ii) $Fe_{Ti}'-V_O^{\bullet\bullet}$ complexes in "hard" PZT, resulting from the acceptor doping [4,57]. While dispersion indicates conductivity by ionic hopping in both undoped and "hard" PZT, we should note that the steep low-frequency dispersion observed here in PZT75/25 and PZT60/40 may be alternatively interpreted by other mechanisms, e.g., Maxwell-Wagner (M-W) relaxation [5]. Independently on the exact nature of the underlying mechanism, however, the data suggests that the low-frequency permittivity of the two PZTs (Fig. 5a,b) is likely affected by conductivity processes, probably involving $V_O^{\bullet\bullet}$. The conjecture is well supported by the observed wide $p(O_2)$-independent region in the conductivity-$p(O_2)$ relationship measured in undoped PZT, which is characteristic of ionic conductivity [58]. In addition, we note that the aging/de-aging behavior of PZT75/25 and PZT60/40, discussed in the previous section, assumes $V_O^{\bullet\bullet}$ motion, which is consistent with the conductivity contribution seen in the permittivity at low frequencies.

The same transition from the $\log(1/\omega)$ dispersion at high driving frequencies to the steep dispersion at low frequencies, as seen in the permittivity (Fig. 5a,b), is also reflected in the direct piezoelectric response (Fig. 5c,d). An important difference with respect to the permittivity, is that the steep $d_{33}$ dispersion in the low frequency range is weaker relative to the log($1/\omega$) dispersion at higher frequencies (see the larger relative change of $\varepsilon_{33}'$ in the low-frequency dispersion regime as compared to $d_{33}$ for both the PZTs). This is somewhat expected considering that the conductivity may directly affect permittivity, e.g., through the relationship $\varepsilon'' \propto \sigma/\omega$ where $\sigma$ is DC conductivity, or through charge hopping where both $\varepsilon'$ and $\varepsilon''$ are affected [59]. In the case of the piezoelectric response, however, the conductivity effect may be indirect, e.g., through the piezoelectric M-W mechanism [60,61] or via domain-wall–pinning-center interactions [62].

We next analyze more closely the two frequency dispersion regimes observed in PZT75/25 and PZT60/40. A way to do this is to examine the area of the *P-E* and *D-σ* subswitching hysteresis loops as a function of the amplitude of the driving electric or stress field [34]. Considering that the hysteresis may originate from a linear (field-independent) or nonlinear (field-dependent) response of the material to the external field, two cases can be separated. The first case is where the hysteresis is directly associated with the nonlinearity, which is described by the Rayleigh law [34]. For an ideal "Rayleigh loop", the area *A* of the hysteresis is given by:

$$A = \frac{4\alpha F_0^3}{3} \qquad (2)$$

where $\alpha$ is the Rayleigh coefficient and $F_0$ is the amplitude of the driving field. In a linear process, where the hysteresis is a consequence of the phase angle between the driving field and the material's response, the area of the hysteresis loop is given by:

$$A = \pi m'' F_0^2 \qquad (3)$$

where $m''$ is the imaginary component of the material's property coefficient, e.g., permittivity or piezoelectric coefficient. Thus, for a nonlinear processes, $A \propto F_0^3$, whereas for linear processes, $A \propto F_0^2$. Depending on the slope of the experimentally determined log($A$) vs. log($F_0$) relationship, we can separate the contribution from a nonlinear hysteretic response (slope 3) from that related to a linear hysteretic response (slope 2).

Figure 6 shows the slope, derived from log($A$) vs. log($F_0$) plots where $F_0$ is either the electric-field amplitude ($E_0$) or stress amplitude ($\sigma_0$), as a function of the driving-field frequency. The data are shown for the dielectric (Fig. 6a) and direct piezoelectric response (Fig. 6b) of PZT75/25 and PZT60/40. In the case of permittivity, we can see that the slope for both PZTs is ~3 in the range 10–100 Hz, and decreases with decreasing frequency, approaching a value closer to 2 (Fig. 6a). This confirms that the response at high frequencies is characterized by nonlinear hysteresis and, thus, irreversible domain-wall displacements (slope 3, Eq. 2), whereas the low-frequency response is linear and dominated by the conductivity (slope 2, Eq. 3). These results are in agreement with the observed transition from the log($1/\omega$) dispersion to the steep dispersion of permittivity with decreasing frequency (Fig. 5a,b).

A similar decreasing trend of the slope with decreasing frequency is observed for the direct piezoelectric response (Fig. 6b). In this case, however, the curves follow a different trend and the slope tends to level off at ~2.5 in the low frequency limit, which would correspond to a contribution from both linear and nonlinear processes to the hysteresis. The PZT60/40 and PZT75/25 ceramics show a slope of ~2.7 and ~3 at 100 Hz, respectively, which means that the piezoelectric response of PZT60/40 consists, in part, of a linear component in the high-frequency range that is not seen in the PZT75/25. A decreasing slope with decreasing frequency was also observed in the converse piezoelectric response (not shown).

Considering that the dielectric and piezoelectric behavior of the tilted PZT75/25 and non-tilted PZT60/40 is qualitatively the same and that the low-frequency dispersion appears in both compositions (Fig. 5, 6), the tilts themselves and their potential elastic interaction with the domain walls [7–9,20,21] cannot explain the appearance of this linear dispersion at low frequencies. In the piezoelectric response, the linear low-frequency behavior may be thus associated either with a reversible domain-wall motion, affected by the motion of $V_O^{\bullet\bullet}$ (conductivity), or M-W mechanism, which is also related to local conductivity in the material, but itself does not necessarily invoke domain-wall motion [60,61].

The link between the conductivity and the linear low-frequency dispersion is further supported by the converse piezoelectric data, shown for three different PZT compositions in Fig. 7a. While the low-frequency dispersion is clearly observed in tilted and non-tilted undoped PZTs (see deviation from $\log(1/\omega)$ dispersion below ~1 Hz for PZT75/25 and PZT60/40 in Fig. 7a), it is not evident in PZT75/25 doped with $Nb_2O_5$ (PZT75/25-Nb). Despite the presence of tilted octahedra in this sample (Fig. 3, PZT75/25-Nb), the doping suppressed the low-frequency dispersion, most likely by reducing the concentration of mobile $V_O^{\bullet\bullet}$, and resulted in a $\log(1/\omega)$ dependence of $d_{33}$ in the whole measured frequency range. The results thus confirm the importance of conductivity and point defects in undoped PZT, regardless of the presence of long-range ordered octahedra tilts in the structure. In addition, this opens up another interesting possibility on the defects-tilts coupling, in which case the motion of $V_O^{\bullet\bullet}$ could be affected by the tilts through elastic interactions, which will be discussed at the end of this section.

The clamping of domain walls by tilts, as suggested earlier [7–10,20,21], and the resulting effect on the frequency dispersion of permittivity and piezoelectric response can be seen by comparing the $\log(1/\omega)$ dispersion in the samples with tilted and non-tilted octahedra and by considering the work of Eitel and Randall [10]. In those studies, they found that the $\alpha$ and $\varepsilon_{init}$ (or $d_{init}$) parameters of the Rayleigh law in the tilted $R3c$ PZT are lower compared to those in the non-tilted $R3m$ PZT; this reduction was attributed to the suppression of the domain-wall motion in the tilted PZT. Since these parameters are frequency dependent [54], one may expect differences in the slope of the $\log(1/\omega)$ dispersion between the tilted and non-tilted PZT, which is, in fact, seen by comparing the data in Fig. 5 and Fig. 7a (the $\log(1/\omega)$ slope in permittivity, direct and converse piezoelectric responses is higher in undoped PZT60/40 than in undoped PZT75/25). We should emphasize, however, that the larger domain-wall contribution in PZT60/40, which was also confirmed by the larger $d_{init}$ (and $\varepsilon_{init}$) and $\alpha$ coefficient (not shown), might be also affected by the close proximity of this composition to the MPB [1,11].

The converse piezoelectric properties, shown in Fig. 7a, were measured by cycling the samples continuously with an alternating electric field. In contrast to the loops measured after several cycles, which did not show significant pinching, we noted that the virgin loops (measured before appreciable cycling) of PZT75/25 and PZT60/40 exhibited a pinched shape (Fig. 7b), in analogy to those measured under switching conditions (Fig. 1, solid curves). In contrast, the Nb-doped PZT75/25 exhibited an open, non-pinched loop in the virgin state (Fig 7b; see also Fig. 4b) with little changes in the loop shape as the sample was cycled. The behavior of PZT75/25 and PZT60/40 upon field cycling provides an opportunity to investigate the possible role of tilts in the de-aging (loop de-pinching) process, which we present next.

Figure 8 shows the $\Delta l$-$E$ loops of tilted PZT75/25 (Fig. 8a) and non-tilted PZT60/40 ceramics (Fig. 8b) before (aged) and after (de-aged) electric-field cycling. In both the cases, the loops experienced depinching by cycling with subswitching fields, which may be attributed to the disordering of the pinning centers [2,33,39], similarly as discussed in the previous section for the case of quenching.

In the next step, we examine the potential role of the octahedra tilts in the disordering of point defect complexes, i.e., we verify whether the tilts may pin elastically the oxygen vacancies, influencing their mobility under field and thus the disordering (de-aging) process. To investigate for such a possibility, we followed the kinetics of the de-aging process in PZT samples with both tilted and non-tilted octahedra. A way to do this is to measure the phase angle of the third harmonic signal of the material's response as a function of cycling time. As demonstrated experimentally in Ref. [33] for the polarization response of "hard" PZT (the principle is generally valid also for the piezoelectric response), the out-of-phase component of the third harmonic polarization signal, $P_3''$, will experience a change in the sign upon the transition from a pinched to a depinched state of the loop. This is because $P_3''$ affects the central part of the loop (around zero E) and, as long as the de-aging process evolves through changes of this central part (e.g., through depinching), $P_3''$ will be strongly affected. Hence, the phase angle of the third harmonic, i.e., $\delta_3 = arc\ tan(P_3''/P_3')$ where $P_3'$ is the in-phase component of the polarization response, may be used to monitor the course of the depinching (de-aging) process and its kinetics.

Figure 8c shows $\delta_3$ of the converse piezoelectric response of PZT75/25 and PZT60/40 as a function of the cycling time. In the aged state, the two ceramics exhibit $\delta_3$ between −180° and −270°, which is consistent with the pinched shape of the loops (Fig. 8a and b, see loops marked as "aged") and with the analysis discussed in Ref. [33]. During subswitching bipolar cycling at 0.2 kV/mm and 1 Hz, $\delta_3$ increases with time, approaching −90° (Fig. 8c, see dashed line), which corresponds to an ideal "Rayleigh-type" loop [34]. This evolution of $\delta_3$ with cycling reflects the de-aging process [33], i.e., the transition from pinched to depinched loops of PZT75/25 and PZT60/40 (Fig. 8a,b). In contrast, the tilted PZT75/25-Nb showed little aging and, correspondingly, a stable $\delta_3$ with cycling (Fig. 8c), close to that of undoped PZT75/25 and PZT60/40 after 7000 s of cycling. This further confirms that the tilts themselves, present in PZT75/25-Nb, are not responsible for the aging and de-aging processes.

As seen from the $\delta_3$ versus cycling time curves (Fig. 8c), the rate of de-aging of PZT75/25 and PZT60/40 is quite similar, however, the tilted PZT75/25 de-ages faster in the first ~3500 s of field cycling. Though the origin of this difference in the de-aging process of the two ceramics is not known, and might be related to a different concentration of pinning centers in the two ceramics, the result rather excludes the possibility that the long-range ordered tilts reduce the $V_O^{\bullet\bullet}$ motion. This is because, in such a case, we would expect a slower de-aging (disordering) in the tilted PZT75/25. The results of the de-aging by subswitching field cycling are thus qualitatively consistent with the de-aging by means of quenching, presented in the previous section.

## IV. SUMMARY AND CONCLUSIONS

We investigated the influence of charged defects on the properties of undoped, rhombohedral PZT both with long-range tilted or non-tilted oxygen octahedra tilts. Based on the analyses of the macroscopic switching behavior and weak-field dielectric and piezoelectric properties, we summarize the results as follows:

1.) Both the tilted *R3c* and non-tilted *R3m* PZT (Fig. 3) exhibit strongly pinched hysteresis loops at switching (Fig. 1) and subswitching (Fig. 7b) driving fields. Pinched loops indicate that the domain walls are pinned; this means that upon release of the applied electric field, a strong restoring force, related to the presence of pinning centers, tends to re-establish the initial (zero-field) domain configuration, resulting in a hysteresis loop with a pinched shape. Early studies on rhombohedral PZT suggest that the domain walls are pinned by local strains associated with oxygen octahedra tilts and that these pinning centers are responsible for the loop pinching. In contrast to these studies, we have shown here that the pinching behavior is rather associated with the pinning effect of charged defects, most probably $V_{Pb}''$ – $V_O^{\bullet\bullet}$ defect complexes. Several experimental data support this conclusion. First, a clear depinching (opening) of the hysteresis loop of PZT75/25 was observed by donor ($Nb^{5+}$) doping (Fig. 4b and Fig. 7b) where this doping is expected to decrease the concentration of pinning centers ($V_O^{\bullet\bullet}$). This depinching occurred despite the PZT preserved the long-range ordering of octahedra tilts after doping (Fig. 3). Second, both the undoped *R3c* and *R3m* PZTs tend to age at room temperature and could be de-aged by quenching (Fig. 1 and Fig. 4a) or continuous electric-field cycling (Fig. 8), a behavior consistent with a transition of the defect complexes from an ordered (aged) state to a disordered (de-aged) state.

2.) The frequency dispersion of permittivity, direct and converse piezoelectric $d_{33}$ coefficients (Fig. 5) is qualitatively the same for *R3c* and *R3m* PZT and it consists of i) a broad, logarithmic dispersion at high frequencies (~0.5–100 Hz), similar to that observed in "soft" (donor-doped) PZT, and ii) a steeper low-frequency dispersion (<~0.5 Hz), reminiscent of that in "hard" (acceptor-doped) PZT. The analysis of the hysteresis loops revealed that the high-frequency dispersion is related to irreversible domain-wall displacements, whereas the low-frequency dispersion could be associated with conductivity-related mechanisms. The effect of charged defects ($V_O^{\bullet\bullet}$) is thus mostly reflected in this low-frequency dispersion of weak-field properties and, in a similar way as the loop pinching, the dispersion disappears when the tilted *R3c* PZT is doped with a donor dopant ($Nb^{5+}$; Fig. 7a), presumably due to a reduction in the concentration of mobile charged defects ($V_O^{\bullet\bullet}$) induced by the donor substitution.

**ACKNOWLEDGEMENTS**

This work was financed by the Slovenian Research Agency (program P2-0105 and project J2-5483) and COST MP0904. We thank Brigita Kmet, Maja Makarovic, Jana Cilensek, Tomaz

Kos and Edi Krajnc for samples preparation and analyses. Goran Dražič is kindly acknowledged for his support in the XRD analysis. We would like to give special thanks to the late Prof. Dr. Marija Kosec for fruitful discussions and her precious advices. The work of Dragan Damjanovic was supported by the Swiss National Foundation Grant No. 200021_159603.

**Figure captions**

FIG. 1. (Color online) *P-E* hysteresis loops of as-sintered and water-quenched (a) PZT75/25 and (b) PZT60/40 ceramics. The inset shows the *P-E* loops of the two quenched samples on the same plot, illustrating the higher coercive field and more pronounced shift of the loop along the electric-field axis of the PZT75/25 sample with respect to the PZT60/40 sample (marked with a circle and an arrow in the inset).

FIG. 2. (Color online) (a) Calculated XRD patterns for *R3c* and *R3m* rhombohedral PZT. The arrows in the *R3c* pattern indicate the additional low-intensity peaks appearing in the *R3c* symmetry, including the $(1\bar{3}\bar{1})$ peak (inset), which are absent in the *R3m* symmetry. (b) Evolution of $(1\bar{3}\bar{1})$ peak of *R3c* PZT75/25 ceramics upon heating to 200°C, i.e., above the tilt transition of 125°C [1], and upon cooling to room temperature, as followed by "in-situ" XRD analysis. Schematic of the annealing schedule during the "in-situ" XRD analysis with indicated tilt transition ($T_t$) is shown in the upper graph of panel (b).

FIG. 3. XRD patterns of as-sintered PZT75/25 ceramics, after quenching in water and after additional 14 days of aging at room temperature with no applied external field. The patterns of as-sintered PZT60/40 and PZT75/25-Nb ceramics are also shown for comparison.

FIG. 4. (Color online) *P-E* hysteresis loops of (a) as-sintered, air-quenched and water-quenched PZT75/25 ceramics and (b) as-sintered PZT75/25-Nb ceramics.

FIG. 5. (Color online) (a,b) Dielectric permittivity ($\varepsilon_{33}'$) and dielectric losses ($tan\delta_d$), and (c,d) direct longitudinal piezoelectric coefficient ($d_{33}$) and tangent of piezoelectric phase angle

($tan\delta_p$) as a function of driving-field frequency for poled (a,c) PZT75/25 and (b,d) PZT60/40 ceramics. The driving electric-field amplitude for permittivity measurements was 0.2 kV/mm, whereas the driving stress amplitude for piezoelectric measurements was 1.9 MPa. The dashed line on the graphs indicates the linear $\log(1/\omega)$ dependence of $\varepsilon_{33}'$ and $d_{33}$ where $\omega$ is frequency.

FIG. 6. (Color online) Slope determined from the linear *log-log* relationship between (a) *P-E* hysteresis loop area and electric-field amplitude ($E_0$) and (b) *D-σ* hysteresis loop area and stress amplitude ($\sigma_0$). The slope is plotted as a function of driving frequency for PZT75/25 (full circles) and PZT60/40 (open circles). The dotted lines are drawn as a guide to the eye.

FIG. 7. (Color online) (a) Relative converse piezoelectric $d_{33}$ coefficient as a function of electric-field frequency for PZT75/25, PZT60/40 and PZT75/25-Nb ceramics. The samples were measured at 0.2 kV/mm. The lines indicate the $\log(1/\omega)$ dependence of $d_{33}$. (b) *Δl-E* hysteresis loops of PZT75/25, PZT60/40 and PZT75/25-Nb ceramics measured at 10 Hz and 0.4 kV/mm.

FIG. 8. (Color online) *Δl-E* hysteresis loops measured at 10 Hz and 0.4 kV/mm of (a) PZT75/25 and (b) PZT60/40 ceramics before (aged) and after exposing the material to continuous bipolar electric-field cycling for 30 min at 0.4 kV/mm and 0.1 Hz (de-aged). After aging of the field-cycled sample at room temperature and zero field ($E=0$) for two months the pinched loops re-appeared (see arrows denoting "aging ($E=0$)" on the graphs). The piezoelectric $d_{33}$ coefficients of the aged and de-aged (field cycled) samples are indicated on the graphs. (c) Evolution of the phase angle of the third harmonic of the mechanical displacement signal ($\delta_3$) for aged PZT75/25 and PZT60/40 during bipolar electric field

cycling at 0.2 kV/mm and 1 Hz. The $\delta_3$-versus-time plot for PZT75/25-Nb is added for comparison.

**Figures**

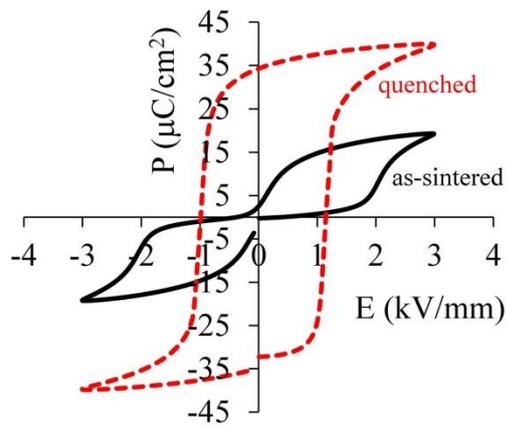 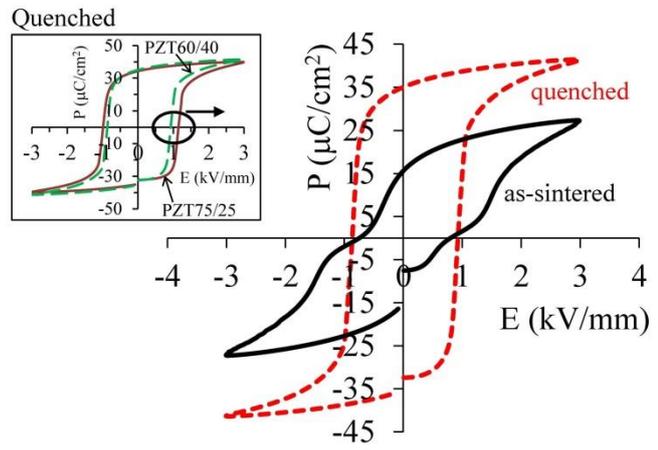

Figure 1

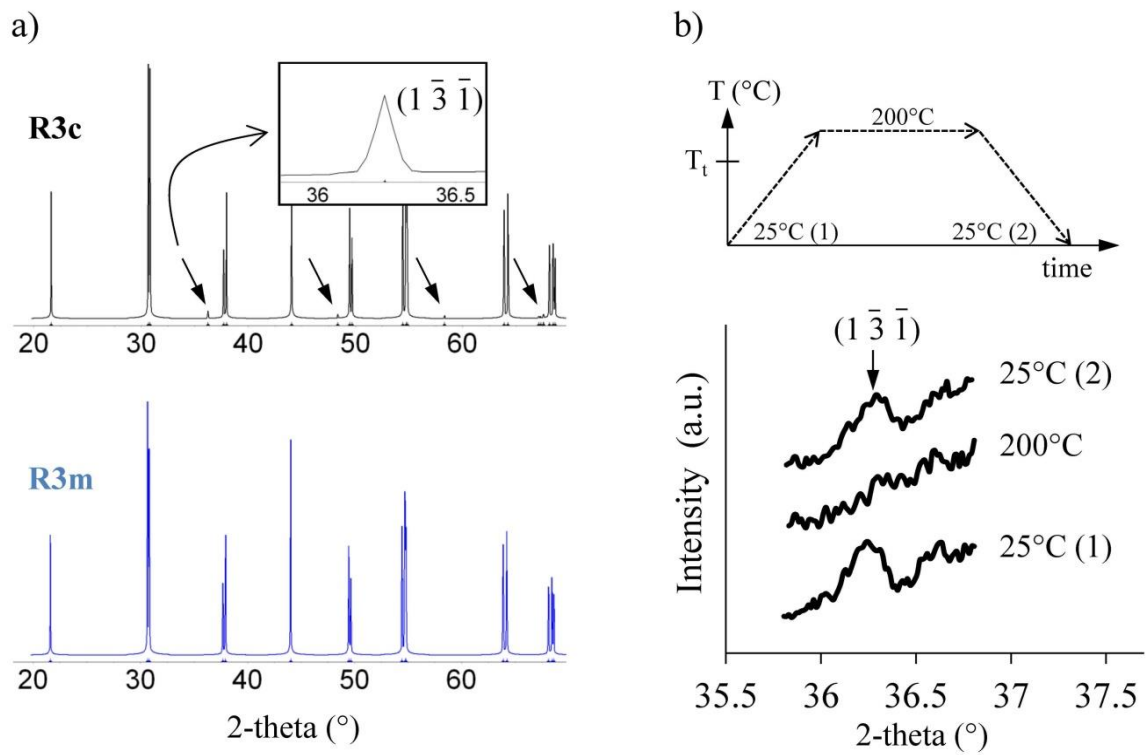

Figure 2

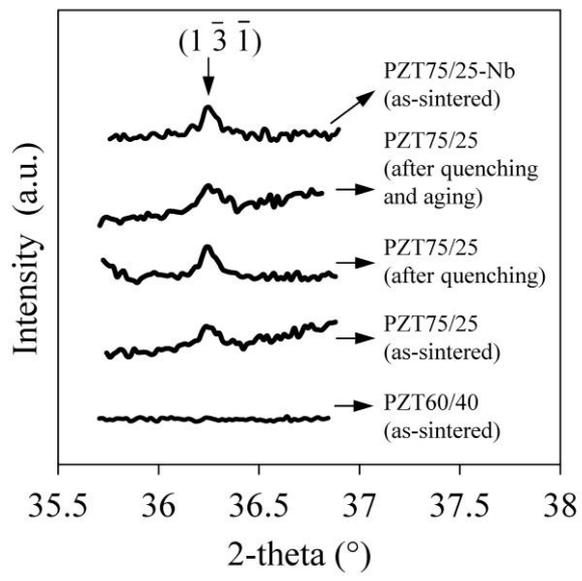

Figure 3

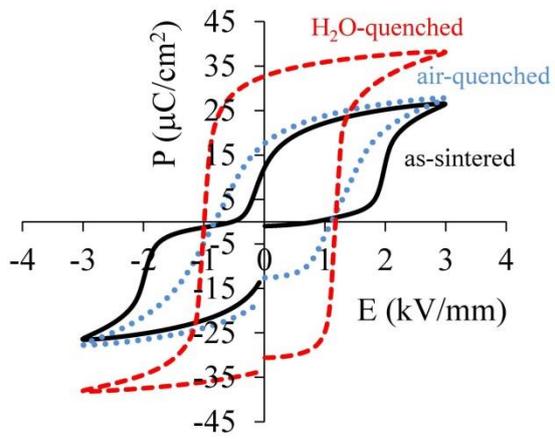 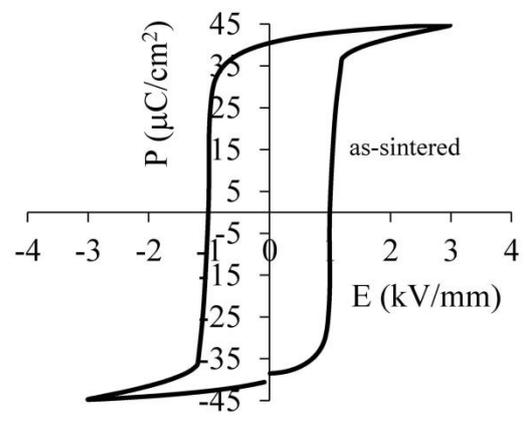

Figure 4

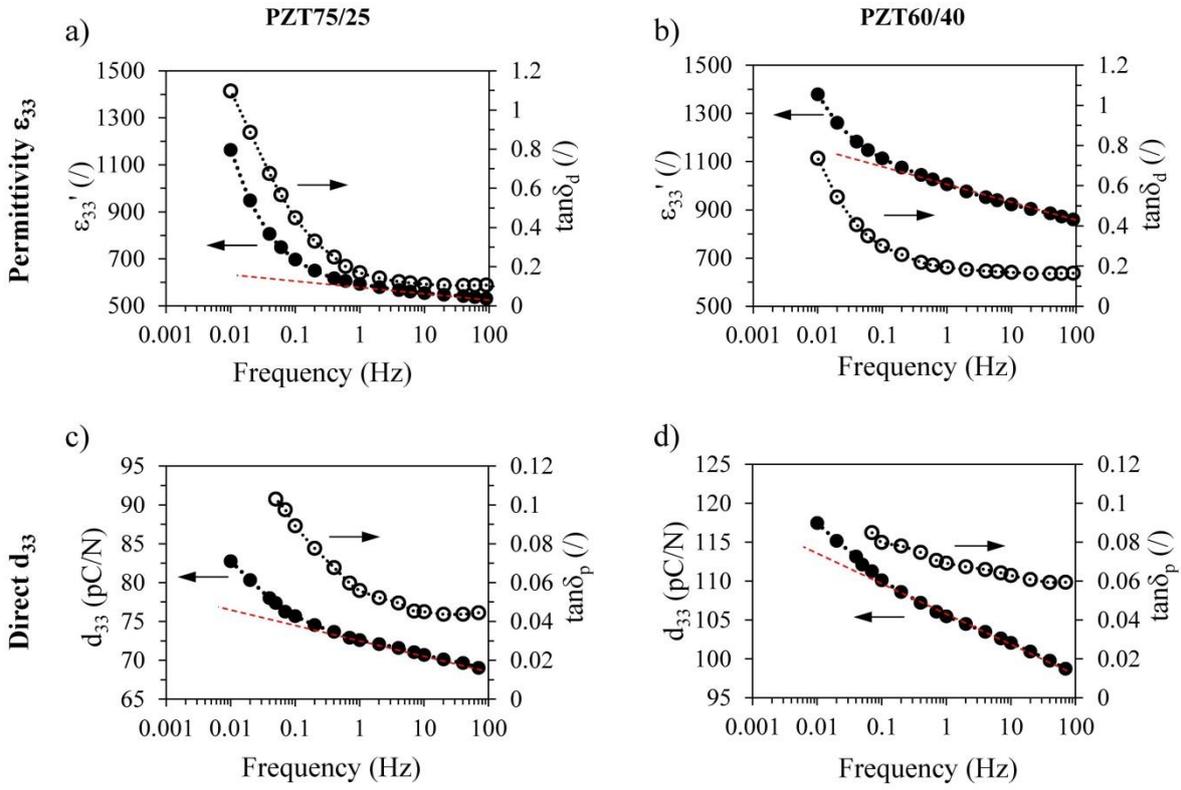

Figure 5

a) Permittivity $\varepsilon_{33}$ 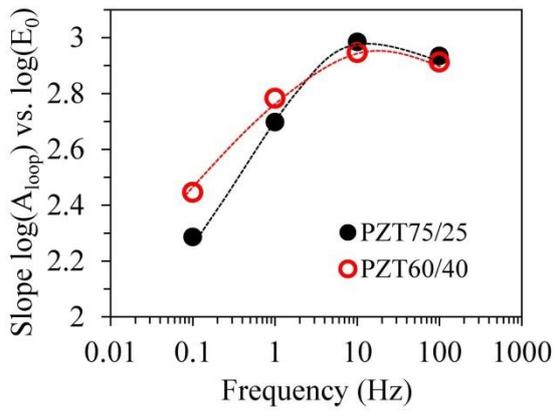

b) Direct $d_{33}$ 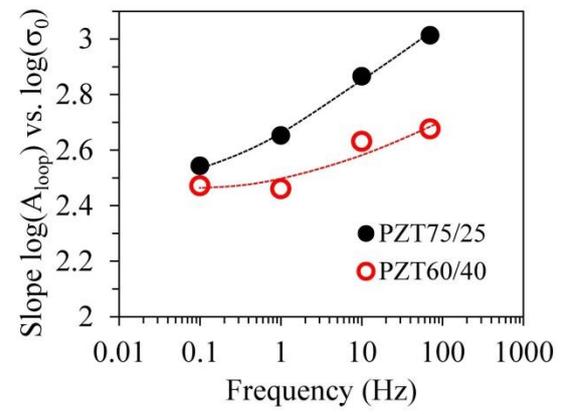

Figure 6

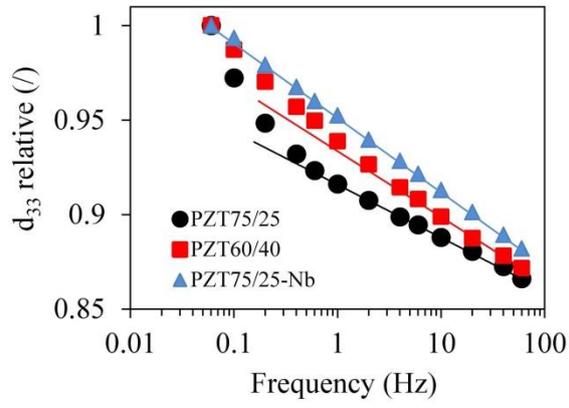 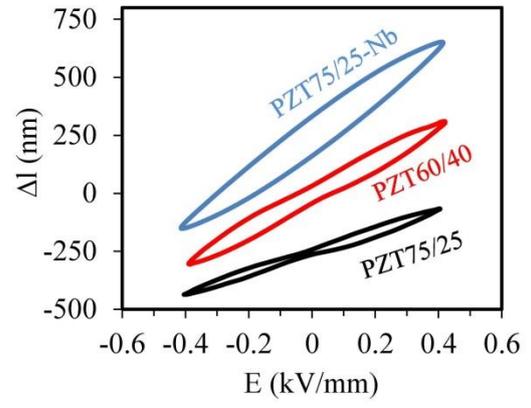

Figure 7

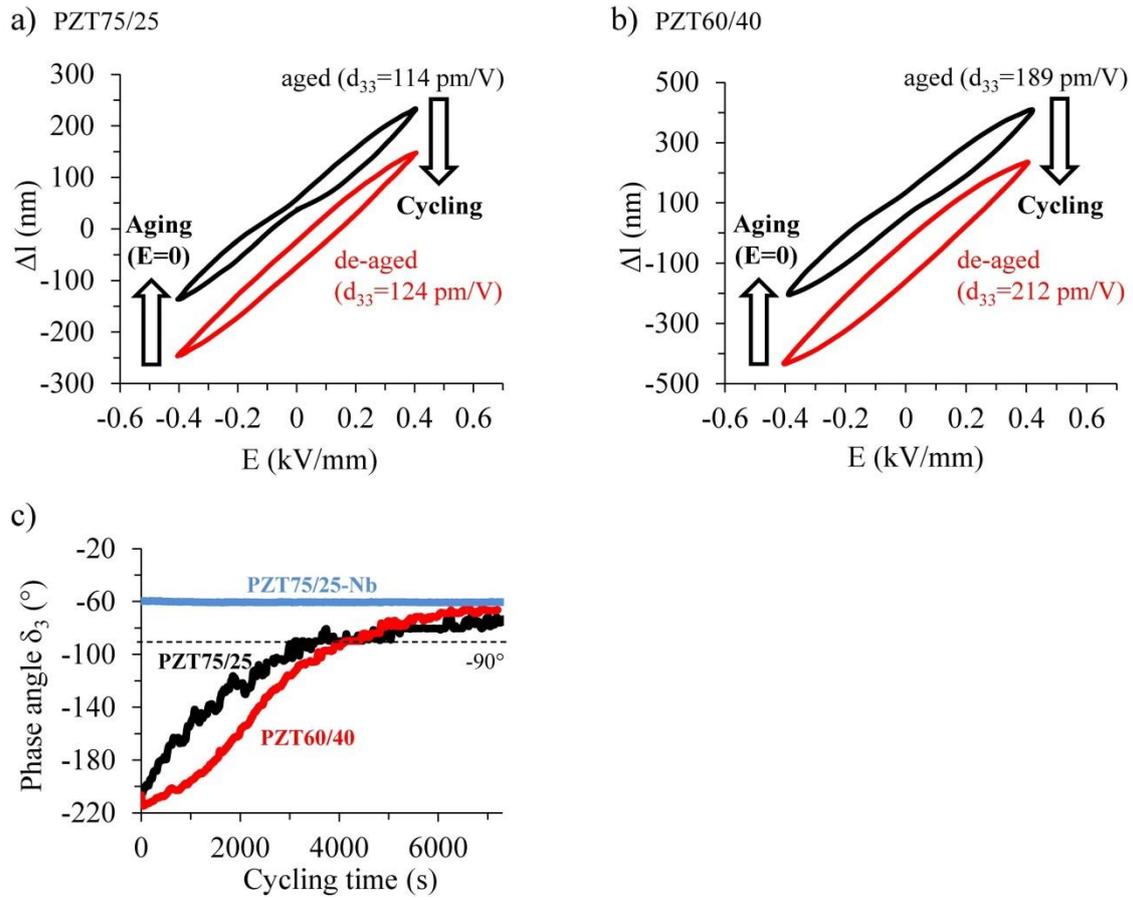

Figure 8